\documentclass[twocolumn,pre,showpacs]{revtex4}
\usepackage{epsfig}
\usepackage{dcolumn}
\usepackage{bm}

\def\be{\begin{equation}}
\def\ee{\end{equation}}

\def\lan{\langle}
\def\ran{\rangle}

\begin{document}
 
 
\title{Glass transition in models with controlled frustration}

\author{Annalisa Fierro}

\affiliation{INFM-Coherentia, Dipartimento di Scienze Fisiche, 
Universit\`a di Napoli ``Federico II'', Complesso Universitario Monte 
Sant'Angelo, Via Cinthia, I-80126, Napoli, Italy}
 
\begin{abstract}
A class of models with self-generated disorder and controlled frustration is
studied. Between the trivial case, where frustration is not present at all, and
the limit case, where frustration is present over every length scale, a
region with local frustration is found where glassy dynamics appears. 
We suggest that in this  region, the mean field model
might undergo a $p$-spin like transition, and increasing  the range of 
frustration, a crossover from a $1$-step replica symmetry breaking 
to a continuous one might be observed.       
\end{abstract} 
\date{\today}

\maketitle       
The study of spin glasses and glass formers has shown that
these systems present similar complex dynamical behaviors. In particular
the dynamical equations of $p$-spin models \cite{pspin}, a generalization
of the spin glass model in mean field, coincide above the dynamical transition 
temperature with
the mode coupling equations for supercooled liquids \cite{MCT}. 
However the connection in finite dimension
between spin glasses and glass formers is not completely clear.
On one hand, spin glasses undergo a
thermodynamic transition at a well defined temperature,
where the nonlinear susceptibility diverges. The systems that show a
transition of this kind, in spite of very different microscopic structures,
have two essential common characteristics: 
The presence of competitive interactions (frustration) and 
quenched disorder. On the other hand, glass formers are a class of systems
where disorder is not originated by some fixed external
variables, but is ``self-generated'' by the particle positions and orientations.
Moreover  there is no sharp thermodynamic transition
characterized by the divergence of a thermodynamic quantity
analogous to the nonlinear susceptibility.                     
In order to clarify the connection and the differences between glasses and
spin glasses, and to investigate the roles of disorder and frustration in the
behaviors observed, in the present paper we study a class of models with 
annealed interactions and controlled frustration.   

We consider a diluted spin glass, the frustrated lattice gas (FLG) 
\cite{varenna},  constituted by 
diffusing particles, and therefore suited to study quantities like the diffusion
coefficient, or the density autocorrelation functions, 
important in the study of liquids. The Hamiltonian of the model is:  
\be
- {\cal H} = J\sum_{\lan ij \ran}
(\epsilon_{ij}S_i S_j - 1)n_in_j +\mu \sum_i n_i,
\label{flg}
\ee
where $S_i=\pm 1$ are Ising spins, $n_i=0,1$ are occupation variables,
and $\epsilon_{ij}=\pm 1$ are ferromagnetic and antiferromagnetic interactions
between nearest neighbor spins.
This model was studied both for quenched \cite{varenna,antonio} and 
annealed interactions \cite{fdc}: 
In the quenched case the interactions, $\epsilon_{ij}$,
are quenched variables randomly distributed with equal probability; in the
annealed case $\epsilon_{ij}$ evolve in time.

In the limit $\mu/T$ goes to infinity all sites are occupied ($n_i\equiv 1$
for each site $i$), and the quenched model reproduces the Ising spin
glass. In the other limit, $T/J=0$, the model, eq. (\ref{flg}), has
properties recalling a ``frustrated'' liquid.
Indeed  the first term of the Hamiltonian implies that two
nearest neighbor sites can be freely occupied only if their spin variables
satisfy the interaction, that is if $\epsilon_{ij}S_iS_j=1$, otherwise
they feel a strong repulsion.  Since in a frustrated loop the spins cannot 
satisfy all the interactions, in this
model particle configurations in which a frustrated loop is fully occupied are
not allowed.  The frustrated loops in the model are the same of the spin glass
model and correspond in the liquid to those loops which, due to geometrical
hindrance, cannot be fully occupied by the particles.   

Here we study a class of annealed FLG models defined  by  the following 
partition function: 
\begin{equation}
{\cal Z}_{an}=\sum^*_{\{ \epsilon \} } \sum_{\{ \sigma\} }
e^{-\beta {\cal H}},
\label{part}
\end{equation}                                                                  
where $\cal H$ is given by eq. (\ref{flg}), the sum $\sum_{\{ \sigma\}}$ is
over all the possible configurations of spin and particles $\{ \sigma\}\equiv
\{ S_i, n_i\}$, and the sum $\sum_{\{ \epsilon \}}^*$
is over all the possible interaction configurations such that
the annealed averages of the frustrated loop number coincide
with the quenched ones for every length of the loops until a maximum value, 
$r_{max}$ ($r_{max} = 0,~4,~6,~8, \dots$ on a cubic lattice).
Varying $r_{max}$ from zero to infinity a class of models with
controlled frustration is obtained. In the limit, $r_{max}=0$, 
frustration is not present at all: the model \cite{fdc} is equivalent to a
lattice gas with a repulsion between nearest neighbors, and without frustration
and correlations between spin, and no thermodynamic transition is present.
In the other limit, $r_{max}$ goes to infinity, frustration is present over 
every length scale, as in the quenched case: We expect that the partition
function, eq. (\ref{part}), coincides with the quenched one, namely the model 
undergoes a spin glass-like transition \cite{antonio}, as shown in
Ref. \cite{PSP} for the Ising spin glass model \cite{nota0}. For intermediate 
values of $r_{max}$ a class of models
with local frustration and self-generated disorder is obtained.

In the present paper the models, eq. (\ref{part}), have been studied for 
$r_{max}=4$ and $r_{max}=6$: In the first case a trivial dynamical behavior is 
observed, with one step
relaxation functions and a smooth increasing of relaxation time as function of
density; in the second case a dynamical behavior very similar to that of 
glass formers is instead found.
We therefore observe that by increasing the degree of frustration the system
moves from a liquid-like to a glassy-like behavior.

The system was simulated using Monte Carlo techniques over a cubic 
lattice of linear size $L=8$.
At the beginning the interactions, $\epsilon_{ij}=\pm 1$, are randomly
distributed with equal probability. At each step of the dynamics an attempt 
to move a particle to a nearest neighbor site (the spin is flipped
with a probability equal to $1/2$) is alternated with an attempt to exchange 
two nearest neighbor interactions: in the limit here considered, $T/J=0$, a
particle can move to a nearest neighbor site only if its spin 
satisfies the interactions with all the new nearest neighbors, and an
interaction can be changed only if at least one of its extreme is empty. 
The frustrated loop numbers of any fixed length until 
$r_{max}$ are independently kept constant during the dynamics.

At a given value of the density, $\rho$, we calculated the two-time 
relaxation function of the self-overlap,
$C(t,t_w)=1/N_p\sum_{i=1}^{L^3}\overline{ S_i(t_w)n_i(t_w)S_i(t)n_i(t)}$, 
where $N_p$ is the number of particles, and the average $\overline{ \ldots} $ 
is done over $8-32$ different realizations of the system.
For values of $t_w$ long enough, the system
reaches a stationary state, where the time translation invariance is
recovered, i.e., $C(t,t_w)=C(t-t_w)$.
In this time region we calculated
the equilibrium relaxation function of the self-overlap:
\be
\lan q\ran (t)={1\over N_p}\sum_{i=1}^{L^3}
\lan S_i(t)n_i(t)S_i(t+t')n_i(t+t')\ran,
\ee
and the dynamical nonlinear susceptibility \cite{franz}:
\be
\chi(t)=N_p(\lan q^2\ran (t)-\lan q\ran^2(t)).
\label{eq_chi}
\ee   
Here  $\lan\cdots\ran$ is the time average on time $t'$. For each density the 
quantities of interest are averaged over $8-32$ different realizations
of the system, and the errors are calculated as the fluctuations  
over this statistical ensemble.

\begin{figure}[ht]
\begin{center}
\mbox{\epsfysize=7cm\epsfbox{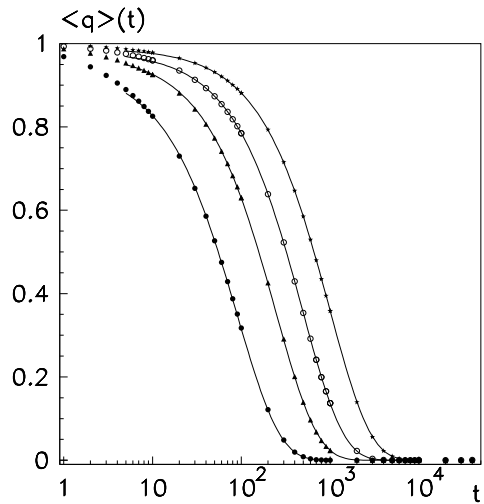}}
\end{center}
\caption{The relaxation functions of the self-overlap, $\lan q\ran (t)$, for 
$r_{max}=4$ at the densities $\rho=~0.63,~0.70,~0.74,~0.76$
(from left to right). The continuous curves in
figure are stretched exponential functions
with $\delta=0.93$.}
\label{fig1r_4}
\end{figure}   
\begin{figure}[ht]
\begin{center}
\mbox{\epsfysize=7cm\epsfbox{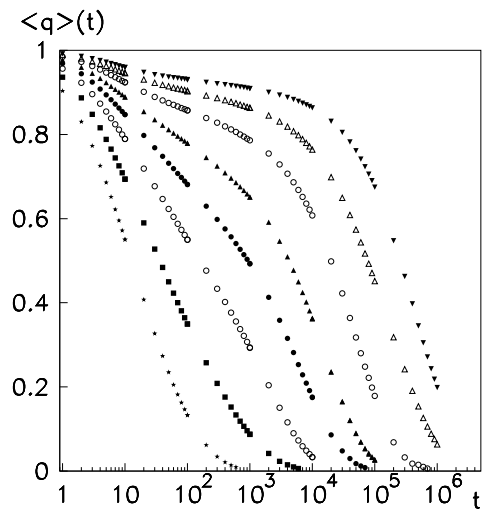}}
\end{center}
\caption{The relaxation functions of the self-overlap, $\lan q\ran (t)$, for
$r_{max}=6$ at the densities $\rho=0.48,~0.54,~0.59,~0.63,~0.66,
~0.71,~0.73$ (from left to right).}
\label{fig_q1}
\end{figure}                                                                    
\begin{figure}[ht]
\begin{center}
\mbox{\epsfysize=7cm\epsfbox{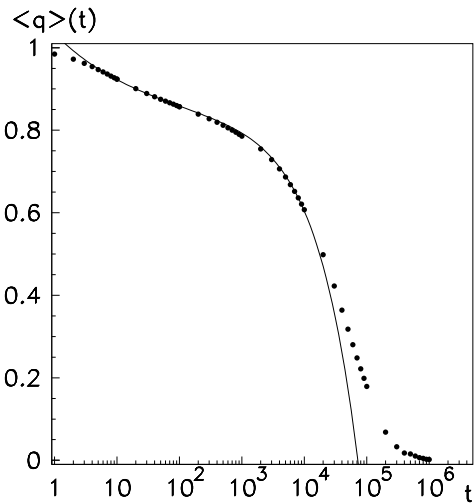}}
\end{center}
\caption{The relaxation functions of the self-overlap, $\lan q\ran (t)$,
for $r_{max}=6$ at the density
$\rho=0.69$. The curve in figure is the $\beta-$ correlator of the MCT
with exponent parameters
$a=0.327$ and $b=0.641$, and plateau height $f_0=0.84$.}
\label{betacorr}
\end{figure}    

We first consider the model with $r_{max}=4$: Since $r_{max}$
equals the loop minimum length, the interactions evolve under the
constraint that the number of frustrated loops of length $4$ is kept constant.
In this case, both $\lan q\ran (t)$ and $\chi(t)$ show a liquid-like behavior 
also at high densities: $\lan q\ran (t)$, plotted in Fig. 
\ref{fig1r_4}, relaxes with a one step decay well fitted by a stretched 
exponential function,
$f(t)=A~ \exp\left\{-\left({t\over\tau}\right)^\delta\right\}$,
with $\delta \simeq 0.93$ not depending on the density; and $\chi(t)$ tends to 
a plateau, which smoothly increases as a function of density.
We suggest that this behavior might be due to the fact that there is only a
few residual frustration on loops of length greater than $4$, and frustration
is too local to originate a slow dynamics. 

A very different behavior is instead observed in the model with 
$r_{max}=6$ (where the interactions evolve under the
constraint that the number of frustrated
loops of length $4$ and $6$ are independently kept constant.)
In Fig. \ref{fig_q1}, $\lan q\ran (t)$ are plotted at different values of the
density. At high density two step decays appear and the curves are well fitted
at intermediate times
by the forms predicted near the dynamical transition by the mode coupling theory
(MCT) \cite{MCT,notaMCT} (see  Fig. \ref{betacorr}).
\begin{figure}[ht]
\begin{center}
\mbox{\epsfysize=7cm\epsfbox{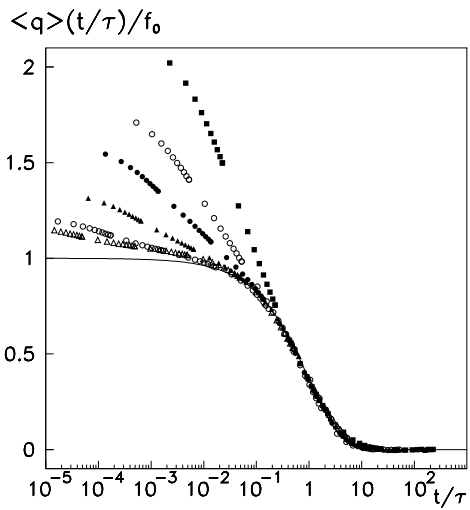}}
\end{center}
\caption{The scaled relaxation functions of the self-overlap, $\lan q\ran
(t/\tau)/f_0$, as a function of the scaled time, $t/\tau$, at 
the densities $\rho=~0.54,~0.59,~0.63,~0.66,~0.71$
(from right to left). The continuous curve is a stretched exponential function
with $\delta=0.71$.}
\label{fig_q3}
\end{figure}
\begin{figure}[ht]
\begin{center}
\mbox{\epsfysize=7cm\epsfbox{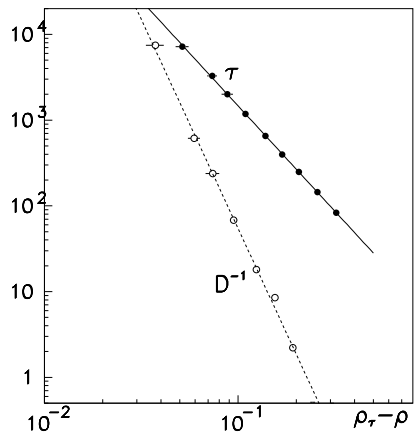}}
\end{center}
\caption{The relaxation time, $\tau$, and the inverse diffusion coefficient,
$D^{-1}$, (the empty circles in figure) are plotted as function of 
$\rho_\tau-\rho$. The curves are the power laws obtained as fitting functions.} 
\label{fig_q4}
\end{figure}                                                                    
In Fig. \ref{fig_q3}, the scaled  relaxation functions of the
self-overlap, $\lan q\ran (t/\tau)/f_0$, are plotted as function of the scaled 
time, $t/\tau$.
At long times the curves collapse onto a single master
function, well fitted by a stretched exponential 
(the continuous curve
in figure), and the relaxation time, $\tau$, 
diverges as a power law, $(\rho_\tau-\rho)^{-\gamma_\tau}$, 
with $\rho_\tau\simeq 0.79\pm 0.02$ and $\gamma_\tau=4.9\pm 0.7$ (see Fig. 
\ref{fig_q4}).
 
\begin{figure}
\begin{center}
\mbox{\epsfysize=7cm\epsfbox{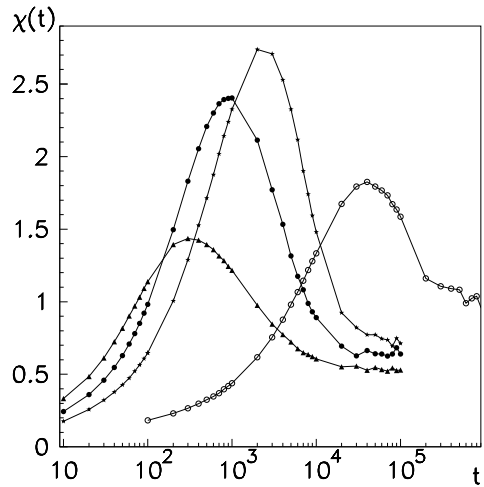}}
\end{center}
\caption{The dynamical nonlinear susceptibility, $\chi(t)$, for $r_{max}=6$ 
at the densities $\rho=0.48,~0.54,~0.59,~0.69$ (from left to right).}
\label{fig3}
\end{figure}                                                                    
The dynamical nonlinear susceptibility, $\chi(t)$, shown in Fig. 
\ref{fig3}, presents a maximum at a time, $t^*$, which we interpret 
as the relaxation time of the interactions: Until $t^*$ the dynamical nonlinear 
susceptibility increases as if the environment were quenched, and 
only for $t > t^*$ the interactions evolve and $\chi(t)$ can decrease to the 
equilibrium value. 
A dynamical nonlinear susceptibility with a maximum is typical of glassy
systems \cite{franz, glotzer}. 
As in Lennard-Jones liquids we found that the value of the maximum,
$\chi(t^*)$, which diverges in the $p$-spin model \cite{franz} as the dynamical 
transition is approached from above, has instead a maximum:
We  suggest that in the present case this behavior might due to the 
fact that the system becomes less and less frustrated as the density increases
\cite{nota}.      

Finally we calculated the particle mean square displacement,
$\lan \Delta r^2\ran (t) = 1/N_p\sum_{i}^{N_p}
\lan ({\bf r}_i(t+t')-{\bf r}_i(t'))^2\ran$,
where ${\bf r}_i(t)$ is the position of the $i-$th particle at the time $t$.
At high density the mean square displacement is not linear at short time, and 
the diffusion coefficient, $D$, is  calculated from the long time regime of
the mean square displacement via the  relation, $D=\lim_{t\rightarrow
\infty} \lan \Delta r^2\ran(t)/6t$.
The diffusion coefficient, shown in Fig. \ref{fig_q4}, is well
fitted by a power law, $(\rho_D-\rho)^{\gamma_D}$, with
$\rho_D=0.80\pm 0.01$ and $\gamma_D=2.46\pm 0.19$.
The critical density, $\rho_D$,  obtained in
this way coincides with the value, $\rho_\tau=0.79\pm 0.02$,
where the relaxation time, $\tau$, diverges; the exponent,
${\gamma_D}$, is instead not consistent with $\gamma_\tau= 4.9\pm0.7$.
 
In conclusions the properties of the annealed models, eq. (\ref{part}), strongly
depend on the value of $r_{max}$. In particular by increasing the degree of 
frustration a crossover from a liquid-like to a glassy-like behavior is 
found. We suggest that the model with $r_{max}=6$, where a dynamical glass 
transition is observed, in mean field might undergo a $p$-spin like 
transition with a $1$-step replica 
symmetry breaking in the spin overlap distribution. Moreover it is reasonable
to expect that, by further increasing the degree of frustration, the annealed 
partition function, eq. (\ref{part}), might tend to the quenched one, where 
a spin glass-like transition \cite{antonio,nota1} and the development of a 
continuous 
replica symmetry breaking in the spin overlap distribution \cite{meanfield} is
found, and a crossover from glassy-like to spin glass-like behavior might
be observed. 
\begin{acknowledgments}
We would like to thank A. Coniglio, E. Del Gado and M. Pica Ciamarra
for many interesting discussions and suggestions. 
This work has been partially supported by EU Network Number
MRTN-CT-2003-504712, MIUR-PRIN 2002, MIUR-FIRB 2002, and INFM-PCI. 
\end{acknowledgments}


\begin{thebibliography}{99}
\bibitem{nota0}
The partition function of the Ising spin glass model is shown to coincide with
that of the annealed model, eq. (\ref{part}), where
${\cal H} =-J \sum_{\lan ij \ran }\epsilon_{ij} S_i S_j$,
the sum $\sum_{\{ \sigma\}}$
is over all the possible spin configurations, and the sum
$\sum_{\{ \epsilon \}}^*$  is over all the possible interaction 
configurations such that the annealed averages of frustrated loop number 
coincide with the quenched ones for every length of the loops.
This was obtained from the observation that the
partition function of the Ising spin glass model for a single realization of
the interactions
depends only on the quantities, $\sum_{\{L_s^{(r)}\}} \prod_{\langle ij\rangle
\in L_s^{(r)}}\epsilon_{ij}$, where the sum is over all the possible loops,
$L_s^{(r)}$, and $s$ labels different loops with the same length, $r$. In the 
present model an analogous proof is not so direct.
\bibitem{notaMCT} 
The $\beta$-correlator predicted by the MCT is given by 
$\Phi(t)=f_c+ g(t/t_\sigma)$, where $g(t/t_\sigma)\propto(t/t_\sigma)^{-a}$
for $t_0\ll t\ll t_\sigma$, and 
$g(t/t_\sigma)\propto-(t/t_\sigma)^b$ for $t_\sigma\ll t\ll\tau_\alpha$.
\bibitem{nota} The annealed model can reach the equilibrium state at density
above the quenched maximum density ($\rho_{max}\simeq 0.68$) only if the
frustration on loops of length above $r_{max}$ is reduced in comparison with the
quenched case. We suggest that for this reason 
the plateau height, $f_0$, which is constant near the glass transition in
the MCT, always increases as function of the density: 
As the frustration length scale decreases also the cage dimension
decreases, and consequently the plateau height increases.
\bibitem{nota1} In the quenched FLG model the relaxation functions of the
spin self-overlap were found to be well fitted by the function
$ct^{-x}\exp(-(t/\tau)^\delta)$, proposed by Olgielski for the Ising spin glass
\cite{olgieski}, and the dynamical nonlinear susceptibility to increase as
function of time to a plateau value, which coincides with the static nonlinear
susceptibility and diverges at the transition \cite{fdc}.
\end{thebibliography}
\end{document}